\documentclass{sig-alternate-05-2015}

\usepackage[utf8x]{inputenc}
\usepackage{listings}
\usepackage{color}
\usepackage{balance}
\usepackage{microtype}
\usepackage{cite}
\usepackage{url}
\usepackage{tabularx}
\usepackage{booktabs}
\usepackage{multirow}
\usepackage{tikz}
\usepackage{amsmath}
\usepackage{ragged2e}
\usepackage{url} 
\usepackage{graphicx}
\usepackage{tikz}
\usepackage{tabularx}
\usepackage{booktabs}
\usepackage{hyperref}
\usepackage{caption}
\usepackage{array}
\usepackage{mdwmath}
\usepackage{mdwtab}
\usepackage{multirow}
\usepackage{longtable}
\usepackage{listings}
\usepackage{color}
\usepackage{balance}

\newcommand{\fakeparagraph}[1]{\smallskip\noindent\textbf{#1.}}

\definecolor{name}{rgb}{0.5,0.5,0.5}
\definecolor{javared}{rgb}{0.6,0,0} 
\definecolor{javagreen}{rgb}{0.25,0.5,0.35} 
\definecolor{javapurple}{rgb}{0.5,0,0.35} 
\definecolor{javadocblue}{rgb}{0.25,0.35,0.75} 

\begin{document}

\setlength{\pdfpageheight}{\paperheight}
\setlength{\pdfpagewidth}{\paperwidth}

\title{Building the Web of Knowledge with Smart IoT Applications\\~(Extended Version)}

\numberofauthors{4} 
%
\author{
Amelie Gyrard\\
       \affaddr{Insight Center for Data Analytics,}\\
			 \affaddr{National University of Galway,}\\
			 \affaddr{Ireland}\\	
       \email{\normalsize{amelie.gyrard@insightcentre.org}}
\alignauthor
Pankesh Patel\\
       \affaddr{ABB Corporate Research,}\\
			 \affaddr{Bangalore}\\
			 \affaddr{India}\\	
       \email{\normalsize{pankesh.patel@in.abb.com}}
\alignauthor
Amit Sheth\\
       \affaddr{Kno.e.sis and Wright State University}\\
			 \affaddr{USA}\\
       \email{\normalsize{amit@knoesis.org}}			
\and
\alignauthor Martin Serrano\\
       \affaddr{Insight Center for Data Analytics,}\\
			 \affaddr{National University of Galway,}\\
			 \affaddr{Ireland}\\	
       \email{\normalsize{martin.serrano@insightcentre.org}}
}

\maketitle
\begin{abstract}
The Internet of Things (IoT) is experiencing fast adoption in the society, from industrial to home applications. The number of deployed sensors and connected devices to the Internet is changing our perspective and the way we understand the world. The development and generation of IoT applications is just starting and they will modify our physical and virtual lives, from how we control remotely appliances at home to how we deal with insurance companies in order to start insurance schemes via smart cards. This massive deployment of IoT devices represents a tremendous economic impact and at the same time offers multiple opportunities. However, the potential of IoT is underexploited and day by day this gap between devices and useful applications is getting bigger. Additionally, the physical and cyber worlds are largely disconnected, requiring a lot of manual efforts to integrate, find, and use information in a meaningful way.

To build a connection between the physical and the virtual, we need a knowledge framework that allow bilateral understandings, devices producing data, information systems managing the data and applications transforming information into meaningful knowledge. The first column in this series in the previous issue of this magazine titled ``Internet of Things to Smart IoT Through Semantic, Cognitive, and Perceptual Computing,'' reviews IoT growth and potential that have energized research and technology development, centered on aspects of Artificial Intelligence to build future intelligent system. This column steps back and demonstrates the benefits of using semantic web technologies to get meaningful knowledge from sensor data to design smart systems.
\end{abstract}
%

\section{Introduction}

We are envisioning a smart IoT system, addressing the key challenges as described below: 
 
The first challenging problem is that devices are not interoperable at any level with each 
other since most of the time technologies differ from one to another. For instance, in 
contemporary IoT applications multiple competing application level protocols such as 
Constrained Application Protocol~(CoAP)\footnote{\url{http://coap.technology/}}, 
Message Queue Telemetry Transport~(MQTT)\footnote{\url{http://mqtt.org/}} and 
Extensible Messaging and Presence Protocol~(XMPP)\footnote{\url{http://xmpp.org/}}
are becoming popular~\cite{DBLP:journals/corr/DesaiSA14}. Each protocol possesses unique 
characteristics and messaging architecture helpful for different types of 
IoT applications. However, a smart IoT application architecture should be 
independent of messaging protocol standards, while also providing integration 
and translation between various popular messaging protocols. 
 
Similarly to proprietary protocols, at the data level, devices do not 
use common terms or vocabulary to describe interoperable IoT data. The 
traditional paradigm of the IoT service model provides unformated data 
names as ``raw'' sensor data. This ``raw'' sensor data does not contain any 
aggregated description (usually representation through semantic annotations) 
and requires specialized knowledge and manual effort in order to build practical 
applications.  
 
Much of the current use of IoT is targeted to a single domain and most
of the times the number of sensors are duplicated unnecessarily. For instance, 
temperature sensors in a building primarily used for a Heating, Ventilating, 
and Air conditioning (HVAC) application. However, values produced by temperature 
sensors could be used in other applications such as fire detection. The primary 
advantage of using common sensors into various applications is that it can reduce 
development, maintenance and deployment costs and promote device reusability. 
To enable cross­domain applications and address interoperability issues, a smart 
IoT system is needed to publish their outputs and to describe device information 
in a well­understood format with added metadata and machine­processable formats, 
thus making devices accessible and usable in multiple applications.  
 
In IoT systems, users are primarily interested in real­world entities 
(such as people, places and things) and their high­level states (e.g., deriving snowfall from temperature 
and precipitation measurements) rather than raw output data produced by sensors attached with these entities. 
To achieve this requirement, a smart IoT system has to provide high­level knowledge that can map sensors to real­world entities and output of raw sensor to high­level states~\cite{swotrommer}. 
 
\subsection{Towards building smart IoT  applications} 
To easily develop IoT applications at a large scale with little or no human intervention, a smart IoT system should leverage semantic web properties, and follow standards. The web of knowledge also plays a relevant role, by defining the rules and mechanisms to associate information in order to produce knowledge. See ``Smart IoT: IoT as a human agent, human extension, and human complement'' in ~\cite{sriram} for the first definition of Smart IoT, which highlights the challenge of interoperating and integrating the data and information.  
 
\begin{figure*}[!ht]
\centering
\includegraphics[width=0.8\textwidth]{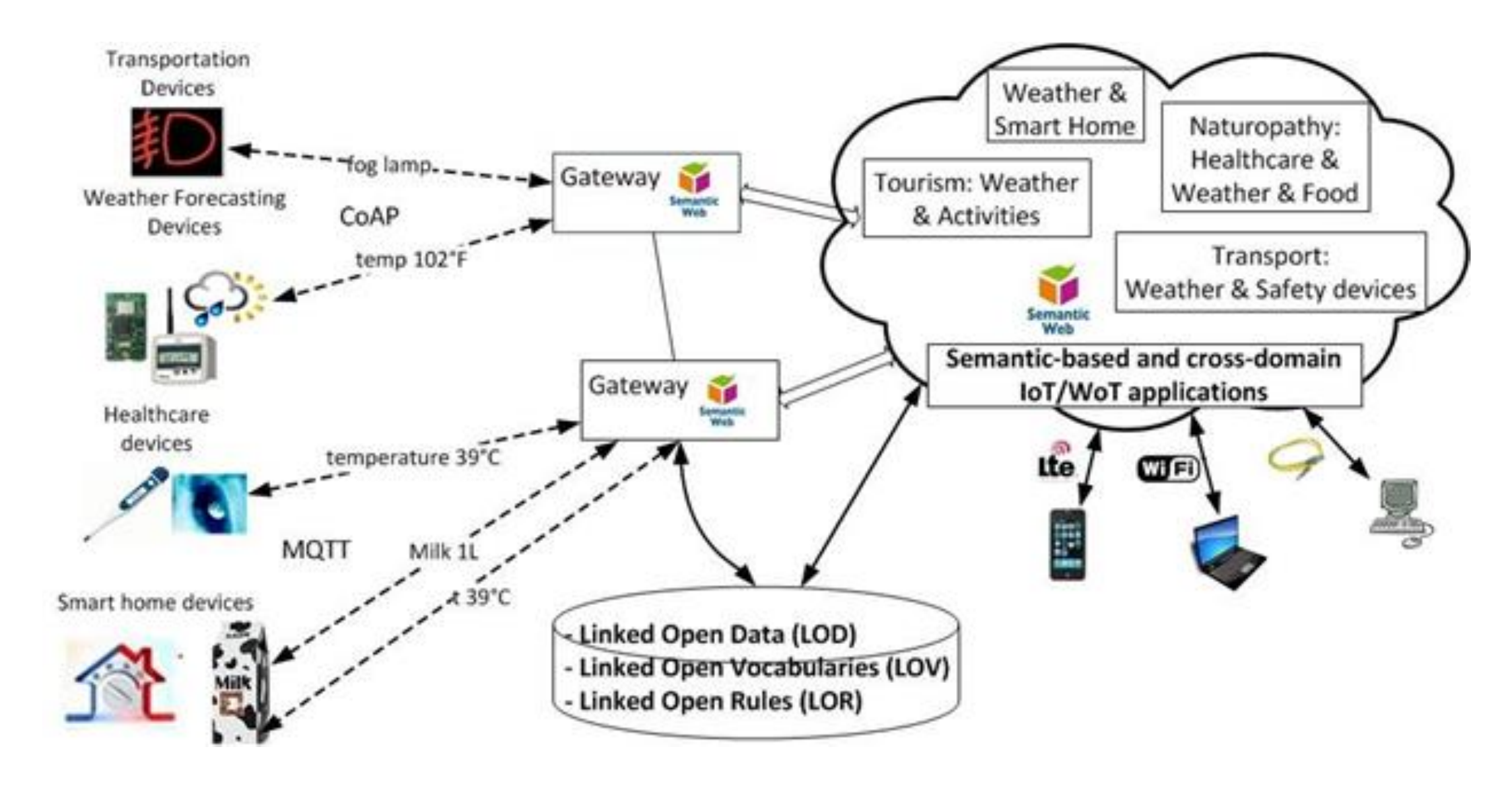}
\caption{Smart IoT System architecture overview~\cite{Gyrard:2013:MAM:2487788.2487945}}
\label{fig:architecture}
\end{figure*}

We are envisioning a smart IoT system that enables good decision making and actions. 
Figure~\ref{fig:architecture} shows an architecture overview of the system inspired by ~\cite{Gyrard:2013:MAM:2487788.2487945}. The architecture largely divided into three layers by their functions: 
 
An example of a smart IoT application is represented below in Figure~\ref{fig:app} taken from the first column in this series (From data to decisions and actions: climbing the data, information, knowledge, and wisdom (DIKW) ladder 
~\cite{sheth2016internet}). The lowest level shows ``150'' which is a blood pressure reading (sensor/device data). The next level shows labeled (semantically annotated) data or information. The third level represents knowledge that is based on the latest NIH guidance used by clinicians, this information represents a medical condition of ``elevated blood pressure''. And yet this knowledge is not actionable--the clinician needs to decide whether this is due to hyperthyroidism or hypertension, which is needed before a proper medication can be prescribed.

\begin{figure}[!ht]
\centering
\includegraphics[width=0.45\textwidth]{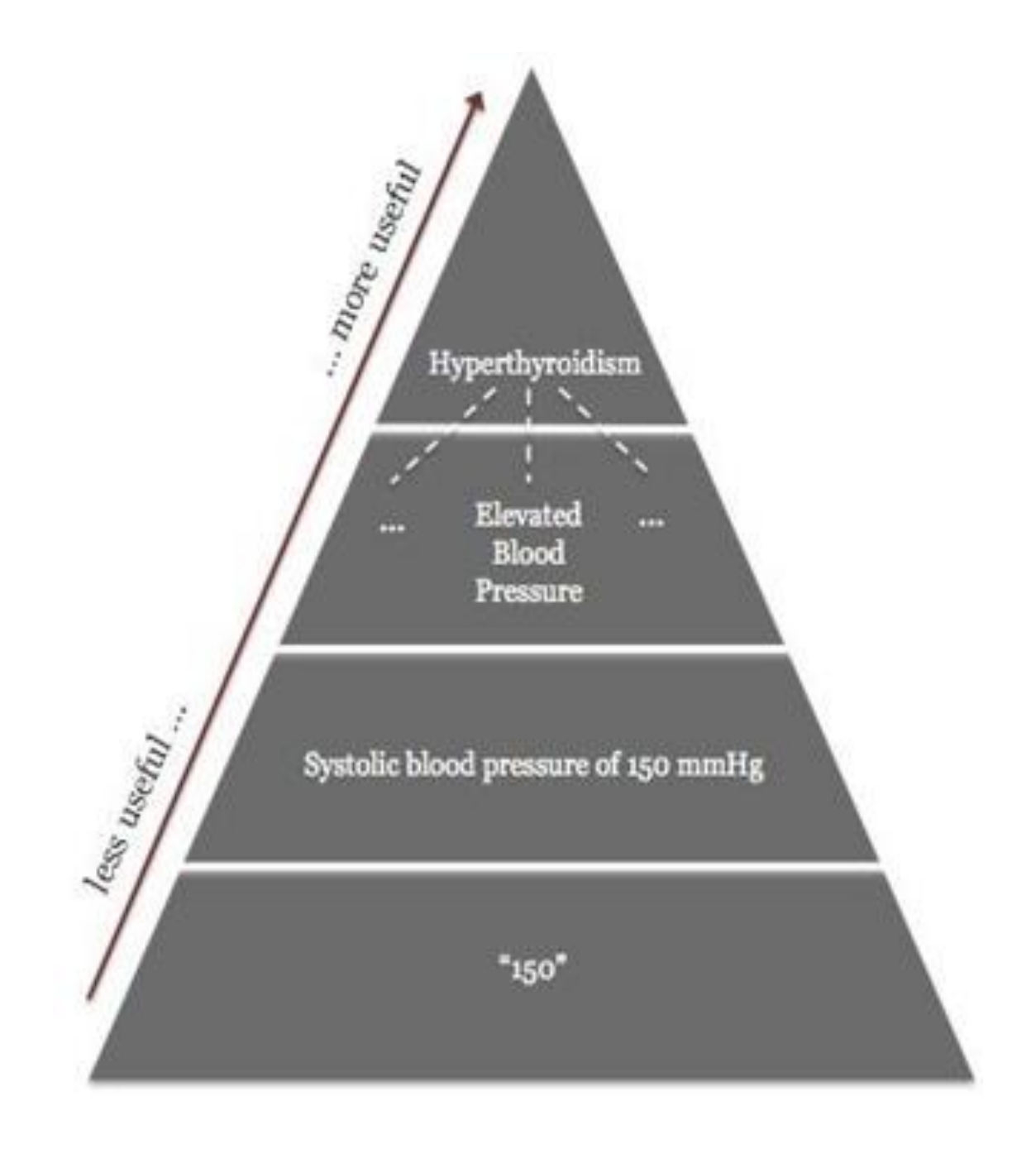}
\caption{Developing a Smart IoT system that supports data to decisions and actions: climbing the data, information, knowledge, and wisdom~(DIKW) ladder~\cite{sheth2016internet} }
\label{fig:app}
\end{figure}

In today's Internet of Things landscape ­ the cyber, virtual and physical worlds are largely disconnected, requiring a lot of manual efforts to integrate, find, and use information in a meaningful way. To realize the application as we discussed above, we are envisioning a smart IoT system that enables good decision making and actions. 
Figure~\ref{fig:architecture} shows an architecture overview of the system inspired by~\cite{Gyrard:2013:MAM:2487788.2487945}. The architecture largely divided into three layers by their functions: 

\begin{enumerate}
	\item \textbf{Accessing things~(Physical)}: This layer is responsible for turning a device such that an application can interact with it. The gateways use device­specific protocols to retrieve data produced by resource­constrained devices. The gateways add semantics to data to unify them, by using semantic web languages (such as RDF, RDFs, OWL) and domain ontologies. 
 
\item \textbf{Deducing new knowledge~(Virtualization)}: The second layer is dedicated to frameworks managing unified data available in standard formats produced by the physical layer. It mainly infers high level knowledge using reasoning engines performed on data and by exploiting the web of knowledge available online.
Such enriched data is provided to the cyber layer to build smart systems, applications and services. 
 
\item \textbf{Composing services~(Cyber)}:  This layer facilitates developers so that they can build large­scale and meaningful IoT applications on top of the virtualization layer. The goal of this level is to drastically reduce the IoT application development, thus enabling rapid prototyping and encourage interoperability of services. 
\end{enumerate}
 
To date, to the best of our knowledge, no concrete and robust technical approaches have been designed to build semantic interoperability for IoT yet. Recently, some semantic interoperability approaches applied to IoT are being designed 
~\cite{barnaghi2012semantics, serrano2015internet,7471300, serrano2013open}. An IoT stack to ensure interoperability has been designed in ~\cite{serrano2015defining}. To define such architecture, semantic interoperability should be provided as explained in ~\cite{sheth1999changing} where it is introduced the idea of an effective approach to bring together metadata, information modeling abstractions and ontologies, as well as the model application domain.  
 
Hereafter, we describe the above three layers, mapping into an example with details that assist developers to design and build smart IoT applications.

\section{Layer~1: Accessing things}
\begin{figure*}[!ht]
\centering
\includegraphics[width=0.7\textwidth]{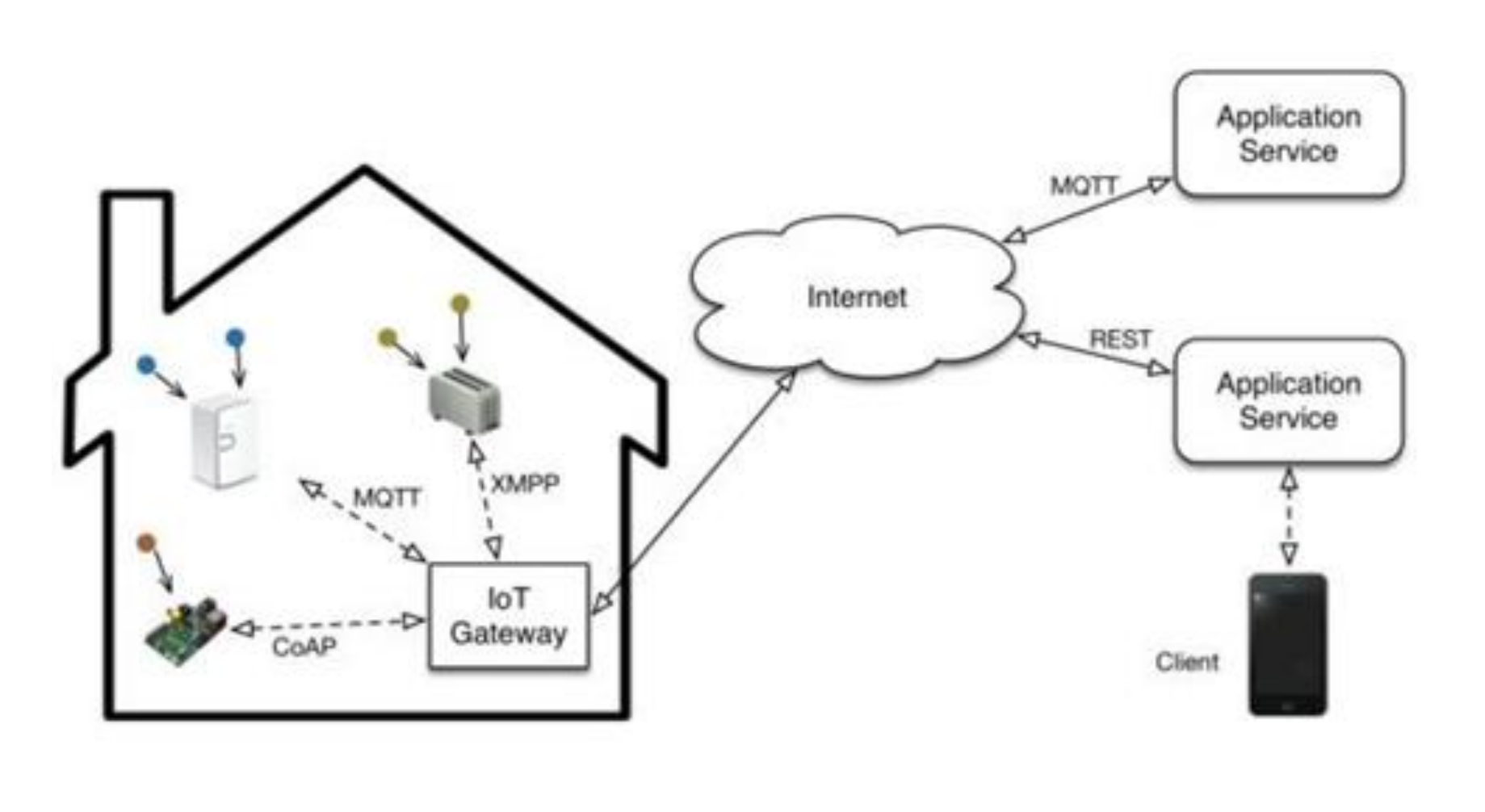}
\caption{Semantic IoT Architecture~\cite{DBLP:journals/corr/DesaiSA14}.}
\label{fig:gateway}
\end{figure*}

The first layer is responsible for turning a device into such that an application can interact with it. The most straightforward way of accessing a device is to expose it and its services directly through APIs. This is applied when a device can support HTTP/Web service (WS) and TCP/IP and can host an HTTP server. However, integrating resource­constrained devices into the Internet is difficult because Internet protocols such as HTTP, TCP/IP are too complex and resource­demanding. To achieve the integration, typically a gateway node is required. To provide interoperability, we can implement necessary technologies at the resource sufficient gateway node. Desai et al. have proposed the concept of ``Semantic Gateway as Service''~(SGS),shown in Figure~\ref{fig:gateway}, that can act as a bridge between resource­constrained devices and IoT application services ~\cite{DBLP:journals/corr/DesaiSA14}. Here, the IoT application services typically collect data from the various gateway nodes and provide user or event specific services using graphics interfaces, notifications or applications.

The SGS architecture broadly provides three functionalities: first, it connects external sink device to the gateway component that supports different protocols such as MQTT, XMPP or CoAP. Second, externally the gateway interfaces cloud services or other SGSs via different protocols such as REST or publish/subscribe. Third, it annotes data acquired from the sink nodes using W3C Semantic Sensor Networks~(SSN)~\cite{compton2012ssn} and domain­specific ontologies before forwarding data to the gateway interface service. The main benefit is that semantic annotation of sensor data by utilizing a standard mechanism and vocabulary can provide interoperability between IoT vertical silos. Semantic Web community has created and optimized standard ontologies for sensor observation, description, discovery and services. By integrating these annotated data and providing Semantic Web enabled messaging interface, a third party service can convert heterogeneous sensor observations to higher level abstractions.  
 
A gateway device such as SGS, discussed above, separates physical level implementation of device to IoT application services. It provides endpoint to IoT application services using a resource interface via REST and publish/subscribe mechanism. The semantic annotation of the sensor data obtained from the gateway assists the IoT services to implement analysis and reasoning algorithms, described in the next layer. 

\section{Layer~2: Deducing new knowledge}  
 
The second layer is dedicated to frameworks managing data and deducing new knowledge. In IoT, most of the time, raw data is just a number~(e.g., 38). Humans implicitly know that data is associated to a specific unit~(e.g., Degree Celsius) and a specific sensor (e.g., thermometer). Smart IoT systems need to interconnect data produced by various sensors to understand the meaning of data to automatically take decisions or to provide suggestions. Dealing with interoperability of heterogeneous data is required to build smarter IoT systems. Data is stored in different files (e.g., CSV, Excel) and structured with different models~(e.g., ontology, schemas). To deal with heterogeneous data, semantic web technologies bring several benefits: (1) unify data, (2) link IoT data to external knowledge bases, (3) explicitly add metadata (i.e., semantic enrichment/enhancement), and (4) deduce new knowledge. Semantic web technologies enable interconnecting knowledge graphs. By interconnecting IoT data with such knowledge graphs (datasets and ontologies used to structure the datasets), IoT systems are becoming smarter. One current form of knowledge graphs that is widely available and useful is ``Linked Open Data'' (LOD)~\cite{bizer2009linked}. Major internet companies such as Facebook and Google are buildding private knowledge graphs based on Schema.org (agreement on common schemas to structure data) that is widely adopted by search engines, in addition to components of LOD such as DBPedia, Wikipedia and/or Wikidata. Such knowledge graphs are build to get access to the information requested more easily and in a automatic way. 
Companies have chosen to use different representation for semantic data, that include unlabeled and labeled graphs, W3C ratified semantic web languages such as RDF and RDFS to explicitly describe the data, and OWL, a language to describe ontologies/vocabularies. 
 
Such semantic technologies provide a basis to later infer high level abstractions from sensor data. Connecting unified semantics­enriched IoT data to the knowledge bases available on the web has a huge potential to build smart systems. For instance, by connecting health measurements to healthcare knowledge bases enable interpreting the raw data itself. For instance, from a body temperature data and by reusing knowledge bases on the web, fever symptom can be deduced. Interconnecting better such knowledge bases, particularly in smart IoT is required.  
 
Currently, open ontology catalogues and dataset catalogues are not interconnecting with each other. New systems are required to link ontologies and datasets, along with the methods to deduce new knowledge from structured data. Taking inspiration from ``Linked Open Data'' and ``Linked Open Vocabularies''~(LOV)~\cite{vandenbussche2015linked} initiatives, ``Linked Open Reasoning'' should be designed and aligned to such initiatives.  
 
\begin{figure*}[!ht]
\centering
\includegraphics[width=0.8\textwidth]{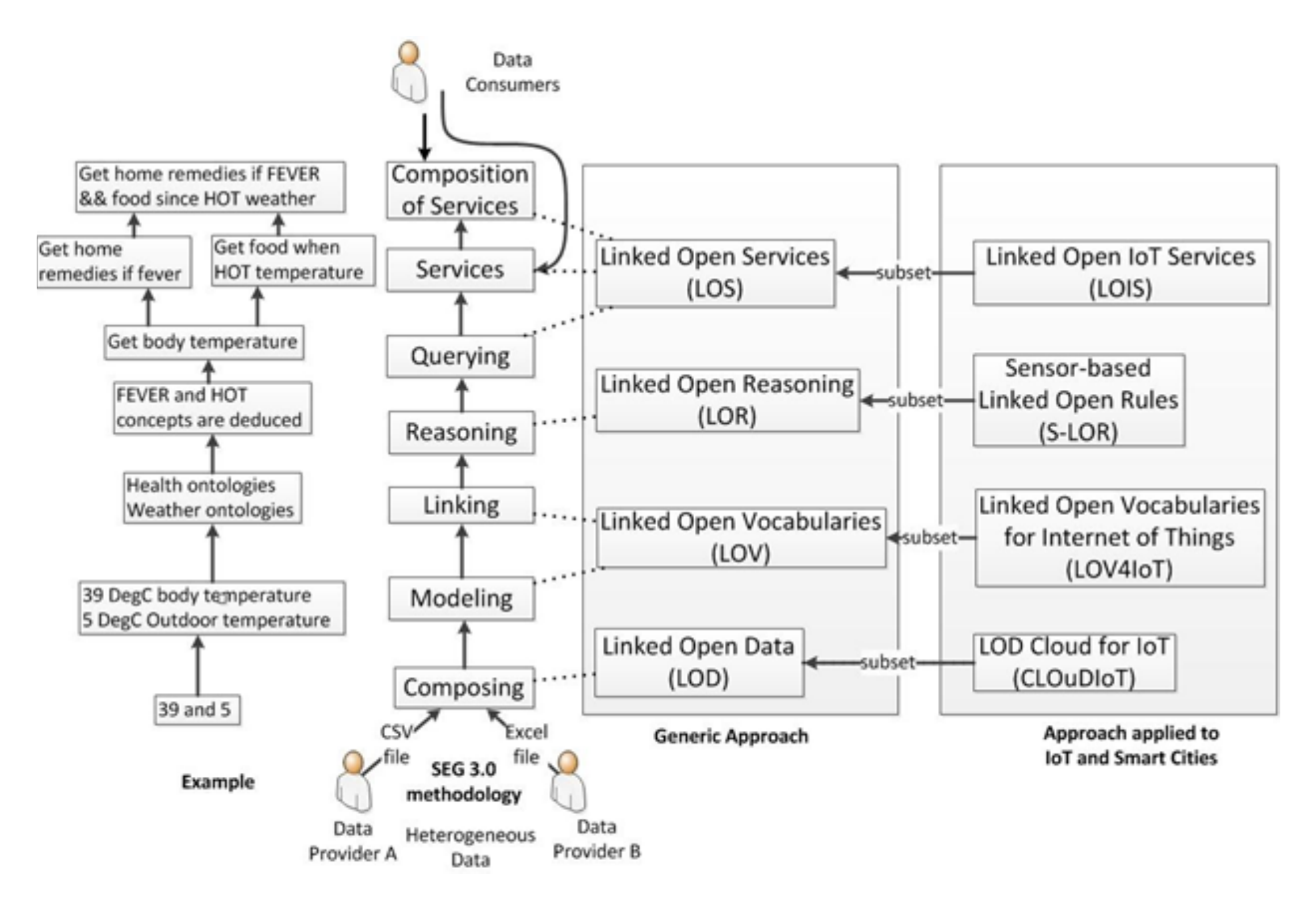}
\caption{The SEG 3.0 methodology ensuring Semantic Interoperability from data providers to data consumers~\cite{7471300}}
\label{fig:SEG}
\end{figure*}

Different processes and steps are required for combining data from heterogeneous sources and for building innovative and interoperable applications. Figure~\ref{fig:SEG} introduces the SEG 3.0 methodology that seeks to meet these requirements and comprises the following steps: (1) composing, (2) modeling, (3) linking, (4) reasoning, (5) querying, (6) services, and (7) composition of services ~\cite{7471300}. The SEG 3.0 methodology encourages the vision to enhance semantic interoperability from data to end­user applications, which is inspired from the 'sharing and reusing' based approach as depicted in Figure~\ref{fig:SEG}.The SEG 3.0 methodology comprises: 

\begin{itemize}
	\item \textbf{Linked Open Data~(LOD)} is an approach to share and reuse data. Previous work regarding `Linked Sensor Data'~\cite{patni2010linked} do not provide any tools for visualizing or navigating through IoT datasets. For this reason, we envision the design of Linked Open Data Cloud for Internet of Things (CLOuDIoT) to share, browse and reuse data produced by sensors.  
 
\item \textbf{Linked Open Vocabularies~(LOV)} is an approach to share and reuse the models/vocabularies/ontologies [Vandenbussche et al. 2015]. To ensure reusability and high quality ontologies, LOV did not reference any ontologies when they do not follow the best practices. Due to this requirement, almost all ontologies for IoT and relevant domain ontologies were not referenced by LOV since IoT community does not yet know the best practices. To overcome this limitation, the Linked Open Vocabularies for Internet of Things (LOV4IoT) has been designed, a dataset of almost 300 ontology­based IoT projects referencing and classifying: (1) IoT applicative domains, (2) sensors used, (3) ontology status (e.g., shared online, best practices followed), (4) reasoning used to infer high level knowledge, and (5) research articles related to the project. This dataset contains a background knowledge required to add value to the data produced by devices. 
 
\item \textbf{Linked Open Reasoning~(LOR)} is an innovative approach to share and reuse the way to interpret data to deduce new information (e.g., machine learning algorithm used, reusing rules already designed by domain experts). Sensor­based Linked Open Rules (S­LOR) is a dataset of interoperable rules (e.g., if then else rules) used to interpret data produced by sensor data~\cite{gyrard2014helping}. Such rules are executed with an inference engine which updates the triple store with additional triples. For example, the execution of the rule ``if the body temperature is greater than 38 degree Celsius than fever'' updates the triple store with the high level knowledge 'fever'. S­LOR is inspired from the idea of 'Linked Rules' which provides a language to interchange semantic rules but not the idea of reusing existing rules.  
 
\item \textbf{Linked Open Services~(LOS)} is an approach to share and reuse the services/applications~\cite{speiser2010towards}
~\cite{serrano2013open}. Composition of services is required to build complex applications. Services can be implemented according to RESTful principles or with the help of semantic web technologies to enhance interoperability 
(e.g., OWL­S). This approach could be extended for designing a set of interoperable services.  
\end{itemize}
 
Sharing and reusing data is insufficient. The entire chain from Linked Open Data~(LOD) to Linked 
Open Services~(LOS) should be shared and reused to enhance interoperability and get meaningful 
knowledge from data. Having this vision in mind, the models, the reasoning and the services 
associated to the data would be interoperable with each other. This entire chain, called SEG 3.0 methodology, has been implemented within the M3 framework~\cite{gyrard2015cross}~\cite{assistingdeveloper} and extended within 
the FIESTA--IoT EU platform\footnote{\url{http://fiesta-iot.eu/}}. M3 enables fast prototyping of IoT applications using 
semantic web technologies to semantically annotate sensor data, deduce new knowledge and combine 
IoT applicative domains. M3 is a semantic engine mainly focused on data interoperability and 
could be used in other EU projects such as FIESTA­IoT, OpenIoT and VITAL. FIESTA­IoT aims to 
achieve interoperability of data, testbeds and experiments by using semantic web 
technologies. One of the component of the FIESTA­IoT project, called ``Experiment--as--a--Service'' demonstrates the proof of concept of the ``Linked Open Services'' approach. Figure~\ref{fig:SEG} highlights an end­to­end scenario 
from raw value~(e.g., 38) to the final application~(e.g., naturopathy to suggest home 
remedies when fever is detected). Such applications can be developed through 
the use of the M3 framework.

\section{Layer~3:~Composition of Services}
Developers can build large­scale and meaningful IoT applications and services on top of devices and IoT data. The goal of this level is to drastically reduce the IoT application development~\cite{patel2016evaluating, patelhal00788366}, thus enabling rapid prototyping. This layer also gets closers to end­user (or domain experts with a limited programming expertise) and enables them to create intelligent applications on top of smart things. In the following, we describe application development approaches for building IoT applications. 
 
\fakeparagraph{General­purpose Programming} Currently, development of IoT is performed at the node­level, by experts of embedded and distributed systems, who are directly concerned with operations of each individual 
device~\cite{patel201562}. For example, developers use general­purpose programming languages (such as JavaScript, C, Java, Android, Python) and target a particular middleware API or node­level service to communicate data. The key advantage of this approach is that it allows the development of efficient systems based on the complete control over individual devices. However, it is unwieldy for IoT applications due to the heterogeneity of systems.  
 
\fakeparagraph{Macroprogramming} It provide abstractions to specify high­level collaborative behaviors, while hiding low­level details such as message passing or state maintenance from stakeholders. A classic example of macro­programming is Node­RED\footnote{\url{http://nodered.org/}}. It is a visual tool for wiring together hardware devices, APIs, and online services. It provides browser­based environment for creating event­driven applications, bridging between physical and cyber services. It contains nodes that can be dragged and dropped into an editor. Each node offers different functionality that can range from a simple debug functionality to accessing sensors via gateways (e.g., Raspberry PI). Macro­programming is viable approach compared to general­purpose programming. However, this approach largely lacks proper software development methodology (e.g., modular design, separation of concerns). That results into a difficult to reuse and platform­dependent design. 
 
\fakeparagraph{Cloud­based Platforms} To improve development effort, the cloud­based platforms provide APIs that provide functions to implement common functionality of IoT applications such as sending and storing data to cloud for data visualization. Moreover, these platforms provide textual and visual programming running on the cloud to write a custom application logic. They provide abstractions to specify high­level collaborative behaviours while hiding low­level details such as message passing. An example of cloud­based approach is IBM­Internet of Things foundations\footnote{\url{ 
https://internetofthings.ibmcloud.com/}}. It is a fully managed and cloud­hosted service that makes it simple to derive value from physical devices. Using abstractions (it is called as recipes), developers can connect devices to the Internet, send sensing data securely to the cloud using the open and lightweight MQTT messaging protocol. From there, developer can leverage various cloud­based services such as dashboard services to visualize and derive insight from the collected data, storage services to store data for historical purpose. 
 
Cloud­based platform is a viable approach compared to the general­purpose programming languages. It reduces development efforts by providing cloud­based APIs to implement common functionality. Second advantage is that because application logic is centrally located, this approach offers the ease deployment and evolution. However, this approach sacrifices direct node­to­node communication. This characteristics restricts developers in­terms of functionality such as “in­network'' aggregation or direct node­to­node communication locally. Third, application logic largely runs on a central cloud, thus an application relies on the availability of cloud provider. So, it may not be suitable for some critical applications.  
 
\fakeparagraph{Model--driven Development~(MDD)} Macroprogramming and cloud­based approach reduce the application development effort. However, they lead to a platform­dependent design. To address this issue, MDD approaches have been proposed. It applies the basic separation of concerns principle both vertically and horizontally. Vertical separation principle reduces the application development complexity by separating the specification Platform Independent Model (PIM) of the system functionality from its platform Platform Specific Model (PSM) such as programming languages. Horizontal separation principle reduces the development complexity by describing a system using different system views, each view describing a certain facet of the system. 
 
An example of MDD approach is IoTSuite~\cite{soukaras2015iotsuite, Chauhan:2016:DFP:2897035.2897039, 7460669}~\cite{patel-icse14, patel-thesis14, patel201562} that aims to make IoT application development easy for developers. It achieves this aim by integrating a set of high­level languages to specify an IoT application. It provides automation techniques to parse the specifications written using these high­level languages and generate platform­specific code. The IoTSuite integrates three high­level languages that abstract platform­specific complexity (i.e., horizontal separation of concerns): (1) Domain Language to describe domain­specific features of an IoT application, (2) Architecture Language to describe application­specific functionality of an IoT application, (3) Deployment Language to describe deployment­specific features consisting information about a physical environment where devices are deployed. The IoTSuite is supported by automation techniques such as code­generator that generates platform­specific code by parsing the specification written using the supported high­level programming languages (i.e., vertical separation of concern). 
 
Ensuring semantic interoperability within IoT is really challenging since physical, virtual and cyber layers deal with heterogeneity of hardware devices, protocols, data and reasoning mechanisms to infer high level knowledge. The SEG 3.0 methodology explained above is mainly focused on data interoperability and is a first step towards building interoperable end­user IoT applications and services. 

\section{Conclusions}
As discussed in the first column in this series (see previous issue), IoT deployment is experimenting a fast adoption because of the foreseen positive impact to change all the aspects of our lives. This is accompanied by corresponding variety or heterogeneity for all aspects of IoT ecosystem, including data, communication and application development frameworks.  The vision of smart IoT, also discussed in the previous column, envisages hiding this heterogeneity and corresponding complexity, while enabling development of applications based on intelligent real-time processing of data produced variety of sensors, along with relevant knowledge.  Semantic methods and Semantic Web standards are key enablers of the three requisite layers of a smart IoT system: accessing things (IoTs), understanding IoT data and deducing new knowledge using background/domain knowledge and structured data, and developing composing services.  This article ends with a review of four approaches to application development in a smart IoT ecosystem.

\balance
\bibliographystyle{abbrv}

\end{document}